\documentclass[twocolumn,aps,prl,floats]{revtex4}
\usepackage{graphicx}
\usepackage{amsmath}
\usepackage{color}

\newcommand{\bk}{{\bf k}}
\newcommand{\br}{{\bf r}}

\begin{document}
\title{Robustness of the nodal $d$-wave spectrum to strongly fluctuating competing order}
\author{W. A. Atkinson$^1$} 
\email{billatkinson@trentu.ca}
\author{J. David Bazak$^1$}
\author{B. M. Andersen$^2$}
\affiliation{{}$^1$Department of Physics and Astronomy, Trent University,  Peterborough Ontario, Canada, K9J 7B8\\
{}$^2$Niels Bohr Institute, University of Copenhagen, DK-2100 Copenhagen, Denmark}
\date{\today}

\begin{abstract}
We resolve an existing controversy between, on the one hand, convincing evidence for the existence of competing order in underdoped cuprates, and, on the other hand, spectroscopic data consistent with a seemingly homogeneous $d$-wave superconductor in the very same compounds. Specifically, we show how short-range fluctuations of the competing order essentially restore the nodal $d$-wave spectrum from the qualitatively distinct folded dispersion resulting from homogeneous coexisting phases. The signatures of the fluctuating competing order can be found mainly in a splitting of the antinodal quasi-particles and, depending of the strength of the competing order, also in small induced nodal gaps as found in recent experiments on underdoped $\mathrm{La_{2-x}Sr_xCuO_4}$.      
\end{abstract}
\maketitle

Many recent experiments point to the prominence of competing phases in underdoped cuprate superconductors, and challenge existing theories for the pseudogap phase.\cite{Tanaka2006,Tacon2006,Lee2007,DoironLeyraud2007,Hufner2008,Ma2008,Kondo2009,Kondo2010,Daou2010,He2011,Wu2011,Ideta2012} Evidence has been found for stripe phases in the La$_2$CuO$_4$\cite{Kivelson2003,Tranquada2007} and YBa$_2$Cu$_3$O$_{6+x}$ families,\cite{Hinkov2008,Haug2010,Wu2011,Ghiringhelli2012} checkerboard and nematic phases in the Bi-based cuprates,\cite{Howald2003,Vershinin2004,Kohsaka2007}  loop-current order in YBa$_2$Cu$_3$O$_{6+x}$,\cite{Fauque2006} and spin glass phases in most of the highly underdoped cuprates.\cite{Panagopoulos2003,Stock2006,Sonier2007}  All of these phases are expected to have clear spectroscopic signatures which, in many cases, involve a Fermi surface reconstruction;  puzzlingly, however, angle resolved photoemission spectroscopy (ARPES), which has become a  powerful probe of the electronic structure of the high temperature superconductors, finds no clear evidence of any reconstruction,\cite{Damascelli2003,Kanigel2007,Chatterjee2010,Yang2011}
even for materials like $\mathrm{La_{2-x}Sr_xCuO_4}$ (LSCO) near $x=1/8$ where signatures of reconstruction  due to stripes are expected to be maximal.\cite{JChang2008,He2011b} A few recent studies report antinodal (AN) particle-hole symmetry breaking in the pseudogap phase of $\mathrm{Pb_{0.55}Bi_{1,55}Sr_{1.6}La_{0.4}CuO_{6+\delta}}$ (Bi2201),\cite{Hashimoto2010,He2011} thus confirming the presence of competing order, but the underlying low-temperature dispersion of the electronic quasi-particles in the nodal  region remains remarkably similar to that of a well-known $d$-wave superconductor. This is still true even in the insulating spin-glass regime at doping levels so low that superconductivity is not yet present.\cite{Chatterjee2010}  In many cases, the competing order appears to be fluctuating rather than static,\cite{Panagopoulos2003,Haug2010} and in this paper we demonstrate that many aspects of the ARPES spectrum, such as the robustness of the nodal $d$-wave spectrum to strong competing order, can be understood if this fact is accounted for.

Motivated by the widespread observation of short-range slow
antiferromagnetic (AF) fluctuations in underdoped cuprates, we focus
on the competition between AF and $d$-wave superconducting (dSC)
order, but we expect our main results may apply also to other
candidates for the competing order. Specifically, we have performed
Monte Carlo (MC) simulations of a two-dimensional dSC in which thermal
fluctuations of both the AF and dSC order parameters are included.
Our main finding is that highly inhomogeneous phases, which are widely
reported experimentally, qualitatively alter the low-energy
quasiparticle dispersion from the homogeneous or weakly disordered
situation.  We find that: (1) unless the competing order is very
strong, the spectrum is indistinguishable from that of the pure dSC,
(2) even for strong competing order the nodal electronic spectrum
is very similar to that of the pure superconductor, and shows no
sign of Fermi surface reconstruction expected from AF order, and (3)
when the competing order is strong, the AN gap has a distinct
energy scale from the nodal region.  These results bridge the
seemingly contradictory findings of simple momentum-space behavior
reported by ARPES versus complex inhomogeneous real-space behavior
reported by local probes.

\begin{figure}[t]
\includegraphics[width=\columnwidth]{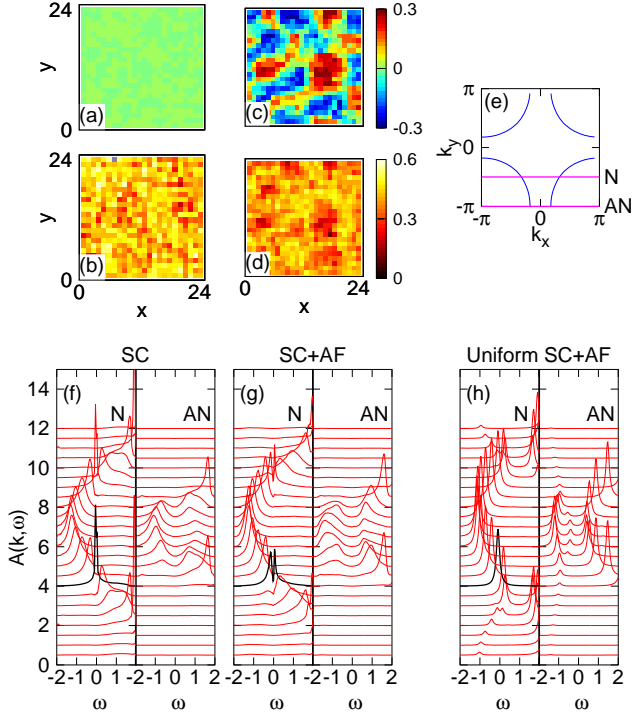}
\caption{(Color online) (a-d) Snapshots of MC simulations for (a,b) a pure dSC and (c,d) a dSC with competing AF order at reduced temperature $\tau=T/T_c=0.3$.
(a,c) The local AF moment $m_Q(\br_i)$ and  (b,d) dSC order parameter are shown for the final step in the MC simulation.  (f-h) The averaged spectral function is shown
along nodal (N) and antinodal (AN) cuts illustrated in (e). The spectra are for (f) dSC only,  (g) dSC+AF, and (h) a homogeneously coexisting dSC and AF. Linewidths are due to thermal fluctuations, except in (h) where they are artificial.  In all cases, the filling is $n\approx 0.88$. Parameters are $V=-1.0$ and (a,b) $U=0$, for which $T_c=0.20$ and (c,d) $U=3.5$, for which $T_c=0.06$.   The Fermi energy is $\varepsilon_F=0$ throughout. In (h) we set $\Delta_d=0.35$ and $Um_Q=3.5\times 0.2$ based on average values in (g).
}
\label{fig:U3.5}
\end{figure}

Our main results are illustrated by Fig.~\ref{fig:U3.5}, where spectra are shown for a pure dSC and for a dSC with strong competing magnetism (dSC+AF). The top panels (a-d) show snapshots, taken from the MC simulation, of the dSC amplitude $\Delta_d(\br_i)$ (b,d) and of the local AF moment $m_Q(\br_i)$ (a,c). The bottom panels (f,g) show the spectral function $A(\bk,\omega)$ at $\bk$-points taken along the nodal and AN lines shown in \ref{fig:U3.5}(e). The spectral function shown in Fig.~\ref{fig:U3.5}(f) [Fig.~\ref{fig:U3.5}(g)] arises from sampling of configurations similar to (a,b) [(c,d)]. For comparison, in Fig.~\ref{fig:U3.5}(h) we also show spectra for the simplified case when dSC and AF order coexist homogeneously.  In this case, we use the root-mean-square values of the AF and dSC order parameters from \ref{fig:U3.5}(g).

The similarity between the nodal spectra for the pure dSC [Fig.~\ref{fig:U3.5}(f)] and dSC+AF [Fig.~\ref{fig:U3.5}(g)] is remarkable, especially in comparison to the uniformly coexisting case [Fig.~\ref{fig:U3.5}(h)].  In the uniform case, the nodal cut exhibits the well-known band back-folding that results in the formation of closed Fermi surface pockets around the AF Brillouin zone boundary. The back-folding is due to the large AF-induced band gap that appears above the Fermi energy.  It is striking that  the band gap is completely absent in \ref{fig:U3.5}(g); instead, the dispersion is almost indistinguishable from that of the pure dSC despite the presence of sizable magnetic moments on nearly all sites evident from Fig.~\ref{fig:U3.5}(c).  The only signature of the competing magnetism is the narrow nodal gap at the Fermi energy in \ref{fig:U3.5}(g) similar to what has been recently found in ARPES studies of underdoped LSCO.\cite{Razzoli2012} We stress that in our model, the nodal gap appears only for large values of $U$, close to the superconductor-insulator transition; at smaller values of $U$, the nodal dispersion remains indistinguishable from the pure dSC case. 

In contrast to the nodal spectrum, the AN spectrum in (g) resembles that of the uniform dSC+AF case in (h):  the single quasiparticle peak at $\omega<\varepsilon_F$ in (f) is split into a pair of peaks in (g) and (h), one of which is upward-dispersing and the other of which is downward-dispersing.  While the splitting is obvious in (h), the peaks are not easily resolved in (g) because the line shapes are thermally broadened, and the lower intensity peak appears as a shoulder.  

The results shown in Fig.~\ref{fig:U3.5}(g) are qualitatively similar to what has been reported at low $T$ for ARPES experiments on Bi2201,\cite{Hashimoto2010,He2011} namely, there is a band splitting at the antinode but no sign of Fermi surface reconstruction near the node.  We emphasize that, while it is possible to model the AN spectrum with a uniform ``finite-$q$'' spin or charge density wave, as in Ref.~\onlinecite{Hashimoto2010}, it is much more difficult to model the spectrum along the entire Fermi surface that way. Indeed, Norman {\em et al.}\cite{Norman2007} have argued against finite-$q$ models of the pseudogap for this reason. Figure~\ref{fig:U3.5}(g) shows that fluctuations qualitatively alter the spectrum in a way that is consistent with experiments.

Figure~\ref{fig:U3.5} is based on MC simulations on an $L\times L$ lattice of electrons coupled to two thermally fluctuating classical fields.  The real field  $h_i$ couples to the local magnetization at site $i$ while the complex field $d_{ij}$ couples to Cooper pairs along bonds connecting nearest-neighbor sites $i$ and $j$.  Purely superconducting models (with $h_i=0$) have been widely used to study phase fluctuations,\cite{Eckl2002,Mayr2005,ValdezBalderas2006,Zhong2011,Banerjee2011}  but relatively little work has focussed on the more complicated problem of competing order.\cite{Alvarez2005,Alvarez2008,Vojta2006}  
 The partition function is 
\begin{equation}
Z= \int D[h d d^\ast] \exp[-\beta \Omega(h,d) ],
\label{eq:Z}
\end{equation}
where
$ \Omega(h,d) = -T \ln \left[ \mbox{Tr } \exp( -\beta \hat H )\right]$, 
 $\mbox{Tr }\ldots$ is a trace over electronic degrees of
freedom, $\int D[h d d^\ast]\ldots$ is a $5L^2$-dimensional
integral over the  fields, and
\begin{eqnarray}
\hat H &=& \sum_{ij\sigma} 
t_{ij} c^\dagger_{i\sigma}c_{j\sigma}  -
\sum_{i\sigma}\sigma h_i \hat n_{i\sigma} +\sum_i \frac{h_i^2}{U} 
\nonumber \\
&& + \sum_{ij}  \left [d_{ij} c^\dagger_{i\uparrow}c^\dagger_{j\downarrow}
+ d_{ij}^\ast c_{j\downarrow}c_{i\uparrow} - \frac{|d_{ij}|^2}{V} \right].
\label{eq:H}
\end{eqnarray}
In Eq.~(\ref{eq:H}), $t_{ij}$ are the hopping matrix elements between
nearest ($t_{ij} = -t=-1$;  $t$ is the unit of energy throughout this work) and next-nearest ($t_{ij} = t^\prime=0.4$) neighbors.  To reduce the number of integration variables, we impose the singlet constraint $d_{ij} = d_{ji}$.   $U$ and $V$ control the size of the AF and SC fluctuations respectively; at $T=0$, the saddle point approximation is exact and gives $h_i = Um(\br_i)$ where $m(\br_i) = \langle \hat n_{i\uparrow}-\hat n_{i\downarrow}\rangle/2$ and $d_{ij} = V\langle c_{j\downarrow} c_{i\uparrow}\rangle$.   Throughout this work, we fix the pairing interaction to be $V=-1.0$ and vary the Coulomb repulsion $U$. The dSC order parameter is $\Delta_d(\br_i) = \sum_j (-1)^{y_i-y_j}\langle c_{j\downarrow}c_{i\uparrow}\rangle$, where $j$ is summed over nearest neighbors of $i$, and the AF moment is $m_Q(\br_i) = (-1)^{x_i+y_i}m(\br_i)$.  The dSC transition occurs at the temperature $T_c$  where the pair correlation $\overline{\Delta_d(\br_i)\Delta^\ast_d(\br_i+{\bf R})} = 0$, with ${\bf R} = \frac 12 (L,L)$.

We use a Metropolis algorithm to evaluate the integral in Eq.~(\ref{eq:Z}).  To reduce noise, all spectra are averaged over 10-20 separate MC simulations;  a typical simulation consists of $2\times 10^4$ sweeps, where one sweep has $L^2$ steps, in which each field is updated.  The MC calculations are computationally demanding because $\hat H$ must be diagonalized at each step; in similar calculations\cite{Alvarez2005} systems were limited to $L\leq 12$ sites (see, however, \cite{Zhong2011} for an alternative approach).  Here, we use first order perturbation theory to estimate updates to $\Omega(h,d)$ at each MC step, with a full diagonalization of $\hat H$ at the beginning of each MC sweep.  We have checked that this does not change the MC-generated distribution of $h_i$ and $d_{ij}$ values, and can study systems up to $L=32$ by this method. 

\begin{figure}[tb]
\includegraphics[width=\columnwidth]{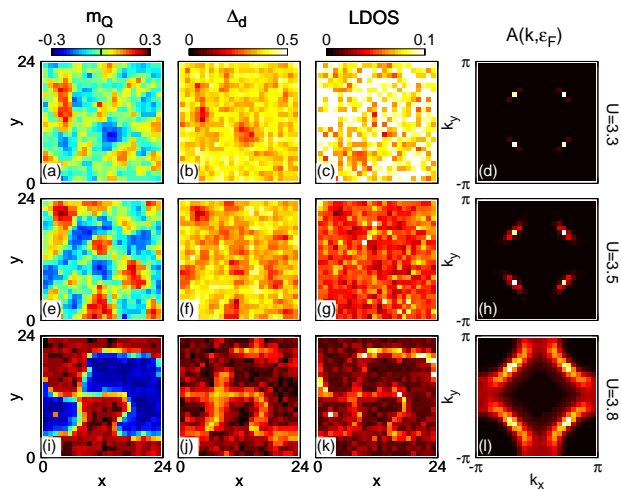}
\caption{(Color online) Results of MC simulations at $T=0.02$ for
  three different $U$ values.  Results are for (a-d) $U=3.3$, (e-h)
  $U=3.5$, and (i-l) $U=3.8$.  First and second columns contain MC
  snapshots of $m_Q(\br_i)$ and $\Delta_d(\br_i)$ respectively; the
  third column contains the LDOS corresponding to the configuration in
  the first two columns.  The final column shows the MC-averaged
  spectral function $A(\bk,\varepsilon_F)$.}
\label{fig:simulations}
\end{figure}
Typical $m_Q$ and $\Delta_d$ configurations sampled by the MC simulations shown in Fig.~\ref{fig:simulations}  illustrate how the model evolves with increasing $U$.  For $U=3.3$ [Fig.~\ref{fig:simulations}(a)], there are large regions where $m_Q(\br_i)$ is small, and small pockets of short-lived AF order.  The dSC order parameter also fluctuates [Fig.~\ref{fig:simulations}(b)], and there is a clear spatial anticorrelation between $\Delta_d(\br_i)$ and $m_Q(\br_i)$. In \ref{fig:simulations}(c), we have plotted the local density of states (LDOS) at the Fermi energy, calculated for the particular configuration of $m_Q(\br_i)$ and $\Delta_d(\br_i)$ shown in \ref{fig:simulations}(a) and \ref{fig:simulations}(b). The LDOS is reduced inside the AF pockets, and is largest in regions where the AF moments are small.  Finally,
in \ref{fig:simulations}(d), we show the MC-averaged spectral function at $\varepsilon_F$.  It is apparent that the nodal quasiparticles are largely unaffected by the fluctuations, so that the low energy spectral weight is concentrated at the four nodal points.

The AF fluctuations are larger when $U=3.5$ [Fig.~\ref{fig:simulations}(e)], and $\Delta_d$ is suppressed in a significant fraction of the sample, although the maximum value of $\Delta_d$ in \ref{fig:simulations}(f) is about the same as in \ref{fig:simulations}(b).   The tendency for the low energy LDOS, shown in \ref{fig:simulations}(g), to be concentrated along AF domain walls is more pronounced than in \ref{fig:simulations}(c), and becomes 
even more pronounced when $U=3.8$, Fig.~\ref{fig:simulations}(k), where there is now phase separation.  In this case, mobile holes lie almost entirely at the boundaries between AF domains; there is some residual pairing of the holes [Fig.~\ref{fig:simulations}(j)], but there is no long-range phase coherence and the system is non-superconducting.  
 
The spectral function at $\varepsilon_F$ for $U=3.8$ shown in Fig.~\ref{fig:simulations}(l) reveals very little of the 
complex real-space structure that emerges as  $U$ increases.  Indeed, the main change is that the nodal points in Fig.~\ref{fig:simulations}(d) evolve into ``Fermi arcs'' with increasing $U$.
That AF fluctuations contribute to arc formation was suggested previously in Ref.~\onlinecite{Alvarez2008}; here we add the observation that this persists even into the nonsuperconducting state.
Indeed, it is remarkable that one recovers the underlying ``bare'' Fermi surface purely from states along the spaghetti-like domain walls.   This situation is reminiscent of the low-energy spectral features studied in the case of disordered stripes,\cite{Salkola1996,Granath2002,Granath2010} which, however, assumed perfect order along one spatial direction leading to a characteristic Fermi surface reconstruction.

\begin{figure}[tb]
\includegraphics[width=0.8\columnwidth]{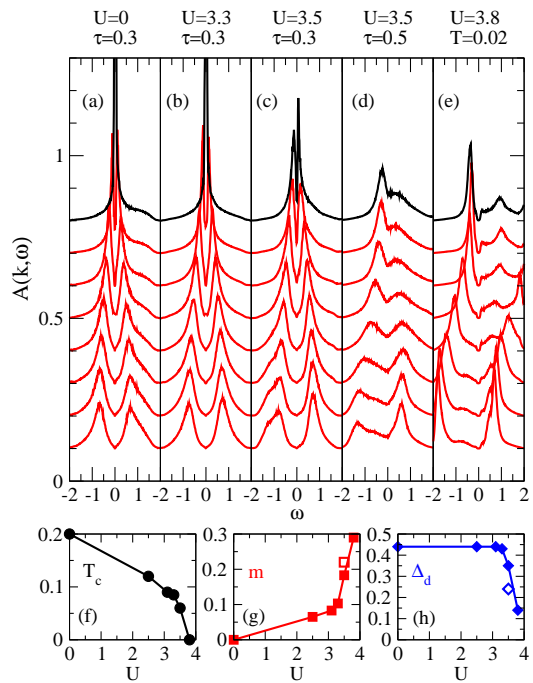}
\caption{(Color online) Spectral function $A(\bk,\omega)$ along the Fermi surface.  Curves (offset for clarity) are taken at $\bk$-points between the nodal point (top curve) to the AN point (bottom curve).   Panels (a)-(e) correspond to different $U$ or reduced temperature  $\tau=T/T_c$, and are arranged in order of increasing average magnetic moment.    $T_c$ values are shown in (f);  root-mean-square averaged (g) $m$ and (h) $\Delta_d$ is shown at $\tau=0.3$ (solid symbols) and $\tau=0.5$ for $U=3.5$ (open symbols).
 }
\label{fig:node}
\end{figure}

Next, we show in Fig.~\ref{fig:node} the progressive evolution of $A(\bk,\omega)$ as the magnetic fluctuations are increased.   The spectra are taken at momenta along the Fermi surface between the nodal and the AN points.  To obtain good momentum resolution, we calculated the spectrum at points that are interpolated between the $L^2$ $\bk$-points of the original simulation as follows: after the MC simulations were completed, we recalculated the spectrum with a complex boundary condition using the previously-generated sequence of $h_i$ and $d_{ij}$ values.  The complex phases were tuned to shift the $\bf k$-values to the Fermi surface (a different phase is required for each $\bk$-point).  
 In panels (a-c) $A(\bk,\omega)$ is shown for increasing $U$ at the same reduced temperature $\tau\equiv T/T_c =0.3$, and the root-mean-square values of $\Delta_d$ and $m$ at this $\tau$ are plotted in \ref{fig:node}(g) and \ref{fig:node}(h), respectively.  In \ref{fig:node}(d), $\tau=0.5$,  which increases $m$ and lowers $\Delta_d$, as shown by the open symbols in \ref{fig:node}(g) and \ref{fig:node}(h).  Finally, in \ref{fig:node}(e), spectra are shown for the phase-separated system at low $T$.

 In agreement with the main conclusions from Fig.~\ref{fig:U3.5}, Fig.~\ref{fig:node}(b) is essentially indistinguishable from Fig.~\ref{fig:node}(a), even though $T_c$ is reduced by a factor of two by AF fluctuations [Fig.~\ref{fig:node}(f)].  Small differences from the $d$-wave spectrum only emerge in \ref{fig:node}(c), by which point $T_c$ is one third of its value in \ref{fig:node}(a).  In \ref{fig:node}(c), a small gap appears at the node and the AN peak below $\varepsilon_F$ splits into two.  These features are more pronounced in \ref{fig:node}(d), where the AF fluctuations are larger.  Finally, in \ref{fig:node}(e), there is a reorganization of the electronic structure into lower and upper magnetic bands. It is the small residual spectral weight near $\varepsilon_F$, coming from the AF domain walls, which generates the Fermi surface in Fig.~\ref{fig:simulations}(l).

\begin{figure}[tb]
\includegraphics[width=0.8\columnwidth]{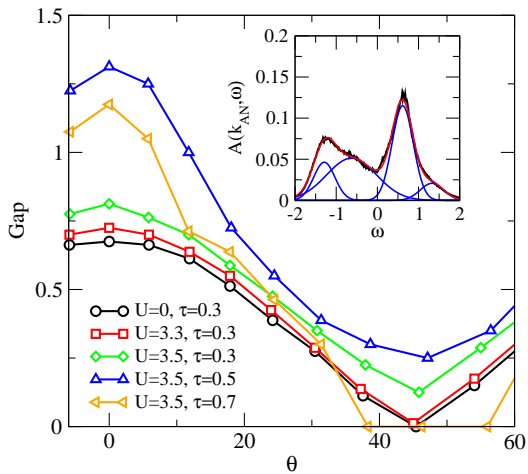}
\caption{(Color online) Spectral gap along the Fermi surface for
  different $U$ and reduced temperature $\tau$.  The angle $\theta$ is
  measured along the Fermi surface, with $\theta=0$ corresponding to
  the antinode and $\theta=45^o$ to the node.  {\em Inset:}
  Decomposition of the antinodal spectrum into gaussians for $U=3.5$,
  $\tau=0.5$. }
\label{fig:fsgap}
\end{figure}

As mentioned above, a recent comprehensive ARPES study of 
LSCO has discovered that the excitation spectrum is gapped along the
entire underlying Fermi surface, and hence displays a nodal gap, in the
highly underdoped regime.\cite{Razzoli2012}  At doping levels
  where the nodal gap is observed, a spin glass is known to coexist
  with superconductivity,\cite{Panagopoulos2003} and in
  Fig.~\ref{fig:fsgap} we show that AF fluctuations can  indeed generate a nodal gap.

Figure~\ref{fig:fsgap} shows how the gap along the Fermi surface
depends on both the magnitude and the correlation length of the AF
fluctuations. To be consistent with experiments, we define the gap
 at a point $\bk$ on the Fermi surface as the position of the
highest-lying peak in $A(\bk,\omega)$ for  $\omega<\varepsilon_F$. 
When $U\leq 3.3$, the gap along the Fermi surface has a dSC-like
structure, and is characterized by a single energy
scale.  When AF correlations are stronger (see e.g. the $U=3.5$,
$\tau=0.3$ curve) the AN gap remains set by the dSC scale, and
a second energy scale emerges in the form of a small gap at the node,
similar to
what was found by Razzoli {\it et al.}\cite{Razzoli2012} In
Fig.~\ref{fig:fsgap} the nodal gap grows with the AF magnitude, but
vanishes at high temperature when the AF correlation length is short
and line broadening wipes it out. A third energy scale emerges at the
AN when AF correlations are strong ($U=3.5$ and $\tau\geq
0.5$). In this case a large gap, with an energy scale distinct from
that near the node, develops.
Similar crossovers between a superconducting nodal gap and
nonsuperconducting AN gap have been widely seen in ARPES
experiments on many underdoped
cuprates;\cite{Tanaka2006,Lee2007,Ma2008,Kondo2009,He2011,Ideta2012}
that this is not reported in LSCO suggests that AF correlations are
not strong enough for a clear AN feature to develop.

We have found that, even when they cannot be resolved by eye,
  two energy scales are always present in the AN spectrum when $U$ is
  large (Fig.~\ref{fig:fsgap} inset).
The inner peak has a $d$-wave $\bk$-space structure, while the outer
peak is tied to the AF fluctuations, although it is not simply related
to the magnetic energy scale $Um \sim 0.7$.  To confirm AF
correlations as the source of the nodal gap in LSCO, our calculations suggest
that one should look for a second energy scale, particularly in the AN
spectrum.  This would distinguish spin glass physics from models
(such as  $d+id$ pairing\cite{Razzoli2012}) in which the gaps
are added in quadrature. 

In conclusion, we have shown that the overall spectrum measured by ARPES experiments is consistent with simple models of competing order if one accounts for the highly inhomogeneous nature of the competing phases. Such models explain the origin of the robust electronic states in momentum space arising from complex real-space environments with substantial amplitudes of the fluctuating competing order.

W.A.A.  and J.D.B. acknowledge support by the Natural Sciences and Engineering Research Council (NSERC) of Canada.  B.M.A. acknowledges support from The Danish Council for Independent Research $|$ Natural Sciences. This work was made possible by the facilities of the Shared Hierarchical Academic Research Computing Network (SHARCNET:www.sharcnet.ca) and Compute/Calcul Canada.


\begin{thebibliography}{44}
\expandafter\ifx\csname natexlab\endcsname\relax\def\natexlab#1{#1}\fi
\expandafter\ifx\csname bibnamefont\endcsname\relax
  \def\bibnamefont#1{#1}\fi
\expandafter\ifx\csname bibfnamefont\endcsname\relax
  \def\bibfnamefont#1{#1}\fi
\expandafter\ifx\csname citenamefont\endcsname\relax
  \def\citenamefont#1{#1}\fi
\expandafter\ifx\csname url\endcsname\relax
  \def\url#1{\texttt{#1}}\fi
\expandafter\ifx\csname urlprefix\endcsname\relax\def\urlprefix{URL }\fi
\providecommand{\bibinfo}[2]{#2}
\providecommand{\eprint}[2][]{\url{#2}}

\bibitem[{\citenamefont{Tanaka et~al.}(2006)\citenamefont{Tanaka, Lee, Lu,
  Fujimori, Fujii, Risdiana, Terasaki, Scalapino, Devereaux, Hussain
  et~al.}}]{Tanaka2006}
\bibinfo{author}{\bibfnamefont{K.}~\bibnamefont{Tanaka}},
  \bibinfo{author}{\bibfnamefont{W.~S.} \bibnamefont{Lee}},
  \bibinfo{author}{\bibfnamefont{D.~H.} \bibnamefont{Lu}},
  \bibinfo{author}{\bibfnamefont{A.}~\bibnamefont{Fujimori}},
  \bibinfo{author}{\bibfnamefont{T.}~\bibnamefont{Fujii}},
  \bibinfo{author}{\bibnamefont{Risdiana}},
  \bibinfo{author}{\bibfnamefont{I.}~\bibnamefont{Terasaki}},
  \bibinfo{author}{\bibfnamefont{D.~J.} \bibnamefont{Scalapino}},
  \bibinfo{author}{\bibfnamefont{T.~P.} \bibnamefont{Devereaux}},
  \bibinfo{author}{\bibfnamefont{Z.}~\bibnamefont{Hussain}},
  \bibnamefont{et~al.}, \bibinfo{journal}{Science}
  \textbf{\bibinfo{volume}{314}}, \bibinfo{pages}{1910} (\bibinfo{year}{2006}).

\bibitem[{\citenamefont{Le~Tacon et~al.}(2006)\citenamefont{Le~Tacon, Sacuto,
  Georges, Kotliar, Gallais, Colson, and Forget}}]{Tacon2006}
\bibinfo{author}{\bibfnamefont{M.}~\bibnamefont{Le~Tacon}},
  \bibinfo{author}{\bibfnamefont{A.}~\bibnamefont{Sacuto}},
  \bibinfo{author}{\bibfnamefont{A.}~\bibnamefont{Georges}},
  \bibinfo{author}{\bibfnamefont{G.}~\bibnamefont{Kotliar}},
  \bibinfo{author}{\bibfnamefont{Y.}~\bibnamefont{Gallais}},
  \bibinfo{author}{\bibfnamefont{D.}~\bibnamefont{Colson}}, \bibnamefont{and}
  \bibinfo{author}{\bibfnamefont{A.}~\bibnamefont{Forget}},
  \bibinfo{journal}{Nature Physics} \textbf{\bibinfo{volume}{2}},
  \bibinfo{pages}{537} (\bibinfo{year}{2006}).

\bibitem[{\citenamefont{Lee et~al.}(2007)\citenamefont{Lee, Vishik, Tanaka, Lu,
  Sasagawa, Nagaosa, Devereaux, Hussain, and Shen}}]{Lee2007}
\bibinfo{author}{\bibfnamefont{W.~S.} \bibnamefont{Lee}},
  \bibinfo{author}{\bibfnamefont{I.~M.} \bibnamefont{Vishik}},
  \bibinfo{author}{\bibfnamefont{K.}~\bibnamefont{Tanaka}},
  \bibinfo{author}{\bibfnamefont{D.~H.} \bibnamefont{Lu}},
  \bibinfo{author}{\bibfnamefont{T.}~\bibnamefont{Sasagawa}},
  \bibinfo{author}{\bibfnamefont{N.}~\bibnamefont{Nagaosa}},
  \bibinfo{author}{\bibfnamefont{T.~P.} \bibnamefont{Devereaux}},
  \bibinfo{author}{\bibfnamefont{Z.}~\bibnamefont{Hussain}}, \bibnamefont{and}
  \bibinfo{author}{\bibfnamefont{Z.~X.} \bibnamefont{Shen}},
  \bibinfo{journal}{Nature} \textbf{\bibinfo{volume}{450}}, \bibinfo{pages}{81}
  (\bibinfo{year}{2007}).

\bibitem[{\citenamefont{Doiron-Leyraud
  et~al.}(2007)\citenamefont{Doiron-Leyraud, Proust, LeBoeuf, Levallois,
  Bonnemaison, Liang, Bonn, Hardy, and Taillefer}}]{DoironLeyraud2007}
\bibinfo{author}{\bibfnamefont{N.}~\bibnamefont{Doiron-Leyraud}},
  \bibinfo{author}{\bibfnamefont{C.}~\bibnamefont{Proust}},
  \bibinfo{author}{\bibfnamefont{D.}~\bibnamefont{LeBoeuf}},
  \bibinfo{author}{\bibfnamefont{J.}~\bibnamefont{Levallois}},
  \bibinfo{author}{\bibfnamefont{J.-B.} \bibnamefont{Bonnemaison}},
  \bibinfo{author}{\bibfnamefont{R.}~\bibnamefont{Liang}},
  \bibinfo{author}{\bibfnamefont{D.~A.} \bibnamefont{Bonn}},
  \bibinfo{author}{\bibfnamefont{W.~N.} \bibnamefont{Hardy}}, \bibnamefont{and}
  \bibinfo{author}{\bibfnamefont{L.}~\bibnamefont{Taillefer}},
  \bibinfo{journal}{Nature} \textbf{\bibinfo{volume}{447}},
  \bibinfo{pages}{565} (\bibinfo{year}{2007}).

\bibitem[{\citenamefont{Hufner et~al.}(2008)\citenamefont{Hufner, Hossain,
  Damascelli, and Sawatzky}}]{Hufner2008}
\bibinfo{author}{\bibfnamefont{S.}~\bibnamefont{Hufner}},
  \bibinfo{author}{\bibfnamefont{M.~A.} \bibnamefont{Hossain}},
  \bibinfo{author}{\bibfnamefont{A.}~\bibnamefont{Damascelli}},
  \bibnamefont{and} \bibinfo{author}{\bibfnamefont{G.~A.}
  \bibnamefont{Sawatzky}}, \bibinfo{journal}{Rep. Prog. Phys.}
  \textbf{\bibinfo{volume}{71}}, \bibinfo{pages}{062501}
  (\bibinfo{year}{2008}).

\bibitem[{\citenamefont{Ma et~al.}(2008)\citenamefont{Ma, Pan, Niestemski,
  Neupane, Xu, Richard, Nakayama, Sato, Takahashi, Luo et~al.}}]{Ma2008}
\bibinfo{author}{\bibfnamefont{J.-H.} \bibnamefont{Ma}},
  \bibinfo{author}{\bibfnamefont{Z.-H.} \bibnamefont{Pan}},
  \bibinfo{author}{\bibfnamefont{F.~C.} \bibnamefont{Niestemski}},
  \bibinfo{author}{\bibfnamefont{M.}~\bibnamefont{Neupane}},
  \bibinfo{author}{\bibfnamefont{Y.-M.} \bibnamefont{Xu}},
  \bibinfo{author}{\bibfnamefont{P.}~\bibnamefont{Richard}},
  \bibinfo{author}{\bibfnamefont{K.}~\bibnamefont{Nakayama}},
  \bibinfo{author}{\bibfnamefont{T.}~\bibnamefont{Sato}},
  \bibinfo{author}{\bibfnamefont{T.}~\bibnamefont{Takahashi}},
  \bibinfo{author}{\bibfnamefont{H.-Q.} \bibnamefont{Luo}},
  \bibnamefont{et~al.}, \bibinfo{journal}{Phys. Rev. Lett.}
  \textbf{\bibinfo{volume}{101}}, \bibinfo{pages}{207002}
  (\bibinfo{year}{2008}).

\bibitem[{\citenamefont{Kondo et~al.}(2009)\citenamefont{Kondo, Khasanov,
  Takeuchi, Schmalian, and Kaminski}}]{Kondo2009}
\bibinfo{author}{\bibfnamefont{T.}~\bibnamefont{Kondo}},
  \bibinfo{author}{\bibfnamefont{R.}~\bibnamefont{Khasanov}},
  \bibinfo{author}{\bibfnamefont{T.}~\bibnamefont{Takeuchi}},
  \bibinfo{author}{\bibfnamefont{J.}~\bibnamefont{Schmalian}},
  \bibnamefont{and} \bibinfo{author}{\bibfnamefont{A.}~\bibnamefont{Kaminski}},
  \bibinfo{journal}{Nature} \textbf{\bibinfo{volume}{457}},
  \bibinfo{pages}{296} (\bibinfo{year}{2009}).

\bibitem[{\citenamefont{Kondo et~al.}(2010)\citenamefont{Kondo, Hamaya,
  Palczewski, Takeuchi, Wen, Xu, Gu, Schmalian, and Kaminski}}]{Kondo2010}
\bibinfo{author}{\bibfnamefont{T.}~\bibnamefont{Kondo}},
  \bibinfo{author}{\bibfnamefont{Y.}~\bibnamefont{Hamaya}},
  \bibinfo{author}{\bibfnamefont{A.~D.} \bibnamefont{Palczewski}},
  \bibinfo{author}{\bibfnamefont{T.}~\bibnamefont{Takeuchi}},
  \bibinfo{author}{\bibfnamefont{J.~S.} \bibnamefont{Wen}},
  \bibinfo{author}{\bibfnamefont{Z.~J.} \bibnamefont{Xu}},
  \bibinfo{author}{\bibfnamefont{G.}~\bibnamefont{Gu}},
  \bibinfo{author}{\bibfnamefont{J.}~\bibnamefont{Schmalian}},
  \bibnamefont{and} \bibinfo{author}{\bibfnamefont{A.}~\bibnamefont{Kaminski}},
  \bibinfo{journal}{Nature Physics} \textbf{\bibinfo{volume}{7}},
  \bibinfo{pages}{21} (\bibinfo{year}{2010}).

\bibitem[{\citenamefont{Daou et~al.}(2010)\citenamefont{Daou, Chang, LeBoeuf,
  Cyr-Choini{\`e}re, Lalibert{\'e}, Doiron-Leyraud, Ramshaw, Liang, Bonn, Hardy
  et~al.}}]{Daou2010}
\bibinfo{author}{\bibfnamefont{R.}~\bibnamefont{Daou}},
  \bibinfo{author}{\bibfnamefont{J.}~\bibnamefont{Chang}},
  \bibinfo{author}{\bibfnamefont{D.}~\bibnamefont{LeBoeuf}},
  \bibinfo{author}{\bibfnamefont{O.}~\bibnamefont{Cyr-Choini{\`e}re}},
  \bibinfo{author}{\bibfnamefont{F.}~\bibnamefont{Lalibert{\'e}}},
  \bibinfo{author}{\bibfnamefont{N.}~\bibnamefont{Doiron-Leyraud}},
  \bibinfo{author}{\bibfnamefont{B.~J.} \bibnamefont{Ramshaw}},
  \bibinfo{author}{\bibfnamefont{R.}~\bibnamefont{Liang}},
  \bibinfo{author}{\bibfnamefont{D.~A.} \bibnamefont{Bonn}},
  \bibinfo{author}{\bibfnamefont{W.~N.} \bibnamefont{Hardy}},
  \bibnamefont{et~al.}, \bibinfo{journal}{Nature}
  \textbf{\bibinfo{volume}{463}}, \bibinfo{pages}{519} (\bibinfo{year}{2010}).

\bibitem[{\citenamefont{He et~al.}(2011{\natexlab{a}})\citenamefont{He,
  Hashimoto, Karapetyan, Koralek, Hinton, Testaud, Nathan, Yoshida, Yao, Tanaka
  et~al.}}]{He2011}
\bibinfo{author}{\bibfnamefont{R.-H.} \bibnamefont{He}},
  \bibinfo{author}{\bibfnamefont{M.}~\bibnamefont{Hashimoto}},
  \bibinfo{author}{\bibfnamefont{H.}~\bibnamefont{Karapetyan}},
  \bibinfo{author}{\bibfnamefont{J.~D.} \bibnamefont{Koralek}},
  \bibinfo{author}{\bibfnamefont{J.~P.} \bibnamefont{Hinton}},
  \bibinfo{author}{\bibfnamefont{J.~P.} \bibnamefont{Testaud}},
  \bibinfo{author}{\bibfnamefont{V.}~\bibnamefont{Nathan}},
  \bibinfo{author}{\bibfnamefont{Y.}~\bibnamefont{Yoshida}},
  \bibinfo{author}{\bibfnamefont{H.}~\bibnamefont{Yao}},
  \bibinfo{author}{\bibfnamefont{K.}~\bibnamefont{Tanaka}},
  \bibnamefont{et~al.}, \bibinfo{journal}{Science}
  \textbf{\bibinfo{volume}{331}}, \bibinfo{pages}{1579}
  (\bibinfo{year}{2011}{\natexlab{a}}).

\bibitem[{\citenamefont{Wu et~al.}(2011)\citenamefont{Wu, Mayaffre, Kr{\"a}mer,
  Horvatic, Berthier, Hardy, Liang, Bonn, and Julien}}]{Wu2011}
\bibinfo{author}{\bibfnamefont{T.}~\bibnamefont{Wu}},
  \bibinfo{author}{\bibfnamefont{H.}~\bibnamefont{Mayaffre}},
  \bibinfo{author}{\bibfnamefont{S.}~\bibnamefont{Kr{\"a}mer}},
  \bibinfo{author}{\bibfnamefont{M.}~\bibnamefont{Horvatic}},
  \bibinfo{author}{\bibfnamefont{C.}~\bibnamefont{Berthier}},
  \bibinfo{author}{\bibfnamefont{W.~N.} \bibnamefont{Hardy}},
  \bibinfo{author}{\bibfnamefont{R.}~\bibnamefont{Liang}},
  \bibinfo{author}{\bibfnamefont{D.~A.} \bibnamefont{Bonn}}, \bibnamefont{and}
  \bibinfo{author}{\bibfnamefont{M.-H.} \bibnamefont{Julien}},
  \bibinfo{journal}{Nature} \textbf{\bibinfo{volume}{477}},
  \bibinfo{pages}{191} (\bibinfo{year}{2011}).

\bibitem[{\citenamefont{Ideta et~al.}(2012)\citenamefont{Ideta, Yoshida,
  Fujimori, Anzai, Fujita, Ino, Arita, Namatame, Taniguchi, Shen
  et~al.}}]{Ideta2012}
\bibinfo{author}{\bibfnamefont{S.-I.} \bibnamefont{Ideta}},
  \bibinfo{author}{\bibfnamefont{T.}~\bibnamefont{Yoshida}},
  \bibinfo{author}{\bibfnamefont{A.}~\bibnamefont{Fujimori}},
  \bibinfo{author}{\bibfnamefont{H.}~\bibnamefont{Anzai}},
  \bibinfo{author}{\bibfnamefont{T.}~\bibnamefont{Fujita}},
  \bibinfo{author}{\bibfnamefont{A.}~\bibnamefont{Ino}},
  \bibinfo{author}{\bibfnamefont{M.}~\bibnamefont{Arita}},
  \bibinfo{author}{\bibfnamefont{H.}~\bibnamefont{Namatame}},
  \bibinfo{author}{\bibfnamefont{M.}~\bibnamefont{Taniguchi}},
  \bibinfo{author}{\bibfnamefont{Z.-X.} \bibnamefont{Shen}},
  \bibnamefont{et~al.}, \bibinfo{journal}{Phys. Rev. B}
  \textbf{\bibinfo{volume}{85}}, \bibinfo{pages}{104515}
  (\bibinfo{year}{2012}).

\bibitem[{\citenamefont{Kivelson et~al.}(2003)\citenamefont{Kivelson, Bindloss,
  Fradkin, Oganesyan, Tranquada, Kapitulnik, and Howald}}]{Kivelson2003}
\bibinfo{author}{\bibfnamefont{S.~A.} \bibnamefont{Kivelson}},
  \bibinfo{author}{\bibfnamefont{I.~P.} \bibnamefont{Bindloss}},
  \bibinfo{author}{\bibfnamefont{E.}~\bibnamefont{Fradkin}},
  \bibinfo{author}{\bibfnamefont{V.}~\bibnamefont{Oganesyan}},
  \bibinfo{author}{\bibfnamefont{J.~M.} \bibnamefont{Tranquada}},
  \bibinfo{author}{\bibfnamefont{A.}~\bibnamefont{Kapitulnik}},
  \bibnamefont{and} \bibinfo{author}{\bibfnamefont{C.}~\bibnamefont{Howald}},
  \bibinfo{journal}{Rev. Mod. Phys.} \textbf{\bibinfo{volume}{75}},
  \bibinfo{pages}{1201} (\bibinfo{year}{2003}).

\bibitem[{\citenamefont{Tranquada}(2007)}]{Tranquada2007}
\bibinfo{author}{\bibfnamefont{J.~M.} \bibnamefont{Tranquada}},
  \bibinfo{journal}{Handbook of High-Temperature Superconductivity Theory and
  Experiment, edited by J. R. Schrieffer}  (\bibinfo{year}{2007}).

\bibitem[{\citenamefont{Hinkov et~al.}(2008)\citenamefont{Hinkov, Haug,
  Fauqu\'e, Bourges, Sidis, Ivanov, Bernhard, Lin, and Keimer}}]{Hinkov2008}
\bibinfo{author}{\bibfnamefont{V.}~\bibnamefont{Hinkov}},
  \bibinfo{author}{\bibfnamefont{D.}~\bibnamefont{Haug}},
  \bibinfo{author}{\bibfnamefont{B.}~\bibnamefont{Fauqu\'e}},
  \bibinfo{author}{\bibfnamefont{P.}~\bibnamefont{Bourges}},
  \bibinfo{author}{\bibfnamefont{Y.}~\bibnamefont{Sidis}},
  \bibinfo{author}{\bibfnamefont{A.}~\bibnamefont{Ivanov}},
  \bibinfo{author}{\bibfnamefont{C.}~\bibnamefont{Bernhard}},
  \bibinfo{author}{\bibfnamefont{C.~T.} \bibnamefont{Lin}}, \bibnamefont{and}
  \bibinfo{author}{\bibfnamefont{B.}~\bibnamefont{Keimer}},
  \bibinfo{journal}{Science} \textbf{\bibinfo{volume}{319}},
  \bibinfo{pages}{597} (\bibinfo{year}{2008}).

\bibitem[{\citenamefont{Haug et~al.}(2010)\citenamefont{Haug, Hinkov, Sidis,
  Bourges, Christensen, Ivanov, Keller, Lin, and Keimer}}]{Haug2010}
\bibinfo{author}{\bibfnamefont{D.}~\bibnamefont{Haug}},
  \bibinfo{author}{\bibfnamefont{V.}~\bibnamefont{Hinkov}},
  \bibinfo{author}{\bibfnamefont{Y.}~\bibnamefont{Sidis}},
  \bibinfo{author}{\bibfnamefont{P.}~\bibnamefont{Bourges}},
  \bibinfo{author}{\bibfnamefont{N.~B.} \bibnamefont{Christensen}},
  \bibinfo{author}{\bibfnamefont{A.}~\bibnamefont{Ivanov}},
  \bibinfo{author}{\bibfnamefont{T.}~\bibnamefont{Keller}},
  \bibinfo{author}{\bibfnamefont{C.~T.} \bibnamefont{Lin}}, \bibnamefont{and}
  \bibinfo{author}{\bibfnamefont{B.}~\bibnamefont{Keimer}},
  \bibinfo{journal}{New J. of Phys.} \textbf{\bibinfo{volume}{12}},
  \bibinfo{pages}{105006} (\bibinfo{year}{2010}).

\bibitem[{\citenamefont{Ghiringhelli and Le~Tacon}(2012)}]{Ghiringhelli2012}
\bibinfo{author}{\bibfnamefont{G.}~\bibnamefont{Ghiringhelli}}
  \bibnamefont{and} \bibinfo{author}{\bibfnamefont{M.}~\bibnamefont{Le~Tacon}},
  \bibinfo{journal}{preprint}  (\bibinfo{year}{2012}).

\bibitem[{\citenamefont{Howald et~al.}(2003)\citenamefont{Howald, Eisaki,
  Kaneko, Greven, and Kapitulnik}}]{Howald2003}
\bibinfo{author}{\bibfnamefont{C.}~\bibnamefont{Howald}},
  \bibinfo{author}{\bibfnamefont{H.}~\bibnamefont{Eisaki}},
  \bibinfo{author}{\bibfnamefont{N.}~\bibnamefont{Kaneko}},
  \bibinfo{author}{\bibfnamefont{M.}~\bibnamefont{Greven}}, \bibnamefont{and}
  \bibinfo{author}{\bibfnamefont{A.}~\bibnamefont{Kapitulnik}},
  \bibinfo{journal}{Phys. Rev. B} \textbf{\bibinfo{volume}{67}},
  \bibinfo{pages}{014533} (\bibinfo{year}{2003}).

\bibitem[{\citenamefont{Vershinin et~al.}(2004)\citenamefont{Vershinin, Misra,
  Ono, Abe, Ando, and Yazdani}}]{Vershinin2004}
\bibinfo{author}{\bibfnamefont{M.}~\bibnamefont{Vershinin}},
  \bibinfo{author}{\bibfnamefont{S.}~\bibnamefont{Misra}},
  \bibinfo{author}{\bibfnamefont{S.}~\bibnamefont{Ono}},
  \bibinfo{author}{\bibfnamefont{Y.}~\bibnamefont{Abe}},
  \bibinfo{author}{\bibfnamefont{Y.}~\bibnamefont{Ando}}, \bibnamefont{and}
  \bibinfo{author}{\bibfnamefont{A.}~\bibnamefont{Yazdani}},
  \bibinfo{journal}{Science} \textbf{\bibinfo{volume}{303}},
  \bibinfo{pages}{1995} (\bibinfo{year}{2004}).

\bibitem[{\citenamefont{Kohsaka et~al.}(2007)\citenamefont{Kohsaka, Taylor,
  Fujita, Schmidt, Lupien, Hanaguri, Azuma, Takano, Eisaki, Takagi
  et~al.}}]{Kohsaka2007}
\bibinfo{author}{\bibfnamefont{Y.}~\bibnamefont{Kohsaka}},
  \bibinfo{author}{\bibfnamefont{C.}~\bibnamefont{Taylor}},
  \bibinfo{author}{\bibfnamefont{K.}~\bibnamefont{Fujita}},
  \bibinfo{author}{\bibfnamefont{A.}~\bibnamefont{Schmidt}},
  \bibinfo{author}{\bibfnamefont{C.}~\bibnamefont{Lupien}},
  \bibinfo{author}{\bibfnamefont{T.}~\bibnamefont{Hanaguri}},
  \bibinfo{author}{\bibfnamefont{M.}~\bibnamefont{Azuma}},
  \bibinfo{author}{\bibfnamefont{M.}~\bibnamefont{Takano}},
  \bibinfo{author}{\bibfnamefont{H.}~\bibnamefont{Eisaki}},
  \bibinfo{author}{\bibfnamefont{H.}~\bibnamefont{Takagi}},
  \bibnamefont{et~al.}, \bibinfo{journal}{Science}
  \textbf{\bibinfo{volume}{315}}, \bibinfo{pages}{1380} (\bibinfo{year}{2007}).

\bibitem[{\citenamefont{Fauqu\'e et~al.}(2006)\citenamefont{Fauqu\'e, Sidis,
  Hinkov, Pailh\`es, Lin, Chaud, and Bourges}}]{Fauque2006}
\bibinfo{author}{\bibfnamefont{B.}~\bibnamefont{Fauqu\'e}},
  \bibinfo{author}{\bibfnamefont{Y.}~\bibnamefont{Sidis}},
  \bibinfo{author}{\bibfnamefont{V.}~\bibnamefont{Hinkov}},
  \bibinfo{author}{\bibfnamefont{S.}~\bibnamefont{Pailh\`es}},
  \bibinfo{author}{\bibfnamefont{C.~T.} \bibnamefont{Lin}},
  \bibinfo{author}{\bibfnamefont{X.}~\bibnamefont{Chaud}}, \bibnamefont{and}
  \bibinfo{author}{\bibfnamefont{P.}~\bibnamefont{Bourges}},
  \bibinfo{journal}{Phys. Rev. Lett.} \textbf{\bibinfo{volume}{96}},
  \bibinfo{pages}{197001} (\bibinfo{year}{2006}).

\bibitem[{\citenamefont{Panagopoulos et~al.}(2003)\citenamefont{Panagopoulos,
  Tallon, Rainford, Cooper, Scott, and Xiang}}]{Panagopoulos2003}
\bibinfo{author}{\bibfnamefont{C.}~\bibnamefont{Panagopoulos}},
  \bibinfo{author}{\bibfnamefont{J.~L.} \bibnamefont{Tallon}},
  \bibinfo{author}{\bibfnamefont{B.~D.} \bibnamefont{Rainford}},
  \bibinfo{author}{\bibfnamefont{J.~R.} \bibnamefont{Cooper}},
  \bibinfo{author}{\bibfnamefont{C.~A.} \bibnamefont{Scott}}, \bibnamefont{and}
  \bibinfo{author}{\bibfnamefont{T.}~\bibnamefont{Xiang}},
  \bibinfo{journal}{Solid State Communications} \textbf{\bibinfo{volume}{126}},
  \bibinfo{pages}{47} (\bibinfo{year}{2003}).

\bibitem[{\citenamefont{Stock et~al.}(2006)\citenamefont{Stock, Buyers, Yamani,
  Broholm, Chung, Tun, Liang, Bonn, Hardy, and Birgeneau}}]{Stock2006}
\bibinfo{author}{\bibfnamefont{C.}~\bibnamefont{Stock}},
  \bibinfo{author}{\bibfnamefont{W.~J.~L.} \bibnamefont{Buyers}},
  \bibinfo{author}{\bibfnamefont{Z.}~\bibnamefont{Yamani}},
  \bibinfo{author}{\bibfnamefont{C.~L.} \bibnamefont{Broholm}},
  \bibinfo{author}{\bibfnamefont{J.-H.} \bibnamefont{Chung}},
  \bibinfo{author}{\bibfnamefont{Z.}~\bibnamefont{Tun}},
  \bibinfo{author}{\bibfnamefont{R.}~\bibnamefont{Liang}},
  \bibinfo{author}{\bibfnamefont{D.}~\bibnamefont{Bonn}},
  \bibinfo{author}{\bibfnamefont{W.~N.} \bibnamefont{Hardy}}, \bibnamefont{and}
  \bibinfo{author}{\bibfnamefont{R.~J.} \bibnamefont{Birgeneau}},
  \bibinfo{journal}{Phys. Rev. B} \textbf{\bibinfo{volume}{73}},
  \bibinfo{pages}{100504} (\bibinfo{year}{2006}).

\bibitem[{\citenamefont{Sonier et~al.}(2007)\citenamefont{Sonier, Callaghan,
  Ando, Kiefl, Brewer, Kaiser, Pacradouni, Sabok-Sayr, Sun, Komiya
  et~al.}}]{Sonier2007}
\bibinfo{author}{\bibfnamefont{J.~E.} \bibnamefont{Sonier}},
  \bibinfo{author}{\bibfnamefont{F.~D.} \bibnamefont{Callaghan}},
  \bibinfo{author}{\bibfnamefont{Y.}~\bibnamefont{Ando}},
  \bibinfo{author}{\bibfnamefont{R.~F.} \bibnamefont{Kiefl}},
  \bibinfo{author}{\bibfnamefont{J.~H.} \bibnamefont{Brewer}},
  \bibinfo{author}{\bibfnamefont{C.~V.} \bibnamefont{Kaiser}},
  \bibinfo{author}{\bibfnamefont{V.}~\bibnamefont{Pacradouni}},
  \bibinfo{author}{\bibfnamefont{S.~A.} \bibnamefont{Sabok-Sayr}},
  \bibinfo{author}{\bibfnamefont{X.~F.} \bibnamefont{Sun}},
  \bibinfo{author}{\bibfnamefont{S.}~\bibnamefont{Komiya}},
  \bibnamefont{et~al.}, \bibinfo{journal}{Phys. Rev. B}
  \textbf{\bibinfo{volume}{76}}, \bibinfo{pages}{064522}
  (\bibinfo{year}{2007}).

\bibitem[{\citenamefont{Damascelli et~al.}(2003)\citenamefont{Damascelli,
  Hussain, and Shen}}]{Damascelli2003}
\bibinfo{author}{\bibfnamefont{A.}~\bibnamefont{Damascelli}},
  \bibinfo{author}{\bibfnamefont{Z.}~\bibnamefont{Hussain}}, \bibnamefont{and}
  \bibinfo{author}{\bibfnamefont{Z.-X.} \bibnamefont{Shen}},
  \bibinfo{journal}{Rev. Mod. Phys.} \textbf{\bibinfo{volume}{75}},
  \bibinfo{pages}{473} (\bibinfo{year}{2003}).

\bibitem[{\citenamefont{Kanigel et~al.}(2007)\citenamefont{Kanigel, Chatterjee,
  Randeria, Norman, Souma, Shi, Li, Raffy, and Campuzano}}]{Kanigel2007}
\bibinfo{author}{\bibfnamefont{A.}~\bibnamefont{Kanigel}},
  \bibinfo{author}{\bibfnamefont{U.}~\bibnamefont{Chatterjee}},
  \bibinfo{author}{\bibfnamefont{M.}~\bibnamefont{Randeria}},
  \bibinfo{author}{\bibfnamefont{M.~R.} \bibnamefont{Norman}},
  \bibinfo{author}{\bibfnamefont{S.}~\bibnamefont{Souma}},
  \bibinfo{author}{\bibfnamefont{M.}~\bibnamefont{Shi}},
  \bibinfo{author}{\bibfnamefont{Z.~Z.} \bibnamefont{Li}},
  \bibinfo{author}{\bibfnamefont{H.}~\bibnamefont{Raffy}}, \bibnamefont{and}
  \bibinfo{author}{\bibfnamefont{J.~C.} \bibnamefont{Campuzano}},
  \bibinfo{journal}{Phys. Rev. Lett.} \textbf{\bibinfo{volume}{99}},
  \bibinfo{pages}{157001} (\bibinfo{year}{2007}).

\bibitem[{\citenamefont{Chatterjee et~al.}(2010)\citenamefont{Chatterjee, Shi,
  Ai, Zhao, Kanigel, Rosenkranz, Raffy, Li, Kadowaki, Hinks
  et~al.}}]{Chatterjee2010}
\bibinfo{author}{\bibfnamefont{U.}~\bibnamefont{Chatterjee}},
  \bibinfo{author}{\bibfnamefont{M.}~\bibnamefont{Shi}},
  \bibinfo{author}{\bibfnamefont{D.}~\bibnamefont{Ai}},
  \bibinfo{author}{\bibfnamefont{J.}~\bibnamefont{Zhao}},
  \bibinfo{author}{\bibfnamefont{A.}~\bibnamefont{Kanigel}},
  \bibinfo{author}{\bibfnamefont{S.}~\bibnamefont{Rosenkranz}},
  \bibinfo{author}{\bibfnamefont{H.}~\bibnamefont{Raffy}},
  \bibinfo{author}{\bibfnamefont{Z.~Z.} \bibnamefont{Li}},
  \bibinfo{author}{\bibfnamefont{K.}~\bibnamefont{Kadowaki}},
  \bibinfo{author}{\bibfnamefont{D.~G.} \bibnamefont{Hinks}},
  \bibnamefont{et~al.}, \bibinfo{journal}{Nature Physics}
  \textbf{\bibinfo{volume}{6}}, \bibinfo{pages}{99} (\bibinfo{year}{2010}).

\bibitem[{\citenamefont{Yang et~al.}(2011)\citenamefont{Yang, Rameau, Pan, Gu,
  Johnson, Claus, Hinks, and Kidd}}]{Yang2011}
\bibinfo{author}{\bibfnamefont{H.-B.} \bibnamefont{Yang}},
  \bibinfo{author}{\bibfnamefont{J.~D.} \bibnamefont{Rameau}},
  \bibinfo{author}{\bibfnamefont{Z.-H.} \bibnamefont{Pan}},
  \bibinfo{author}{\bibfnamefont{G.~D.} \bibnamefont{Gu}},
  \bibinfo{author}{\bibfnamefont{P.~D.} \bibnamefont{Johnson}},
  \bibinfo{author}{\bibfnamefont{H.}~\bibnamefont{Claus}},
  \bibinfo{author}{\bibfnamefont{D.~G.} \bibnamefont{Hinks}}, \bibnamefont{and}
  \bibinfo{author}{\bibfnamefont{T.~E.} \bibnamefont{Kidd}},
  \bibinfo{journal}{Phys. Rev. Lett.} \textbf{\bibinfo{volume}{107}},
  \bibinfo{pages}{047003} (\bibinfo{year}{2011}).

\bibitem[{\citenamefont{Chang et~al.}(2008)\citenamefont{Chang, Sassa,
  Guerrero, M{\aa}nsson, Shi, Pailhés, Bendounan, Mottl, Claesson, Tjernberg
  et~al.}}]{JChang2008}
\bibinfo{author}{\bibfnamefont{J.}~\bibnamefont{Chang}},
  \bibinfo{author}{\bibfnamefont{Y.}~\bibnamefont{Sassa}},
  \bibinfo{author}{\bibfnamefont{S.}~\bibnamefont{Guerrero}},
  \bibinfo{author}{\bibfnamefont{M.}~\bibnamefont{M{\aa}nsson}},
  \bibinfo{author}{\bibfnamefont{M.}~\bibnamefont{Shi}},
  \bibinfo{author}{\bibfnamefont{S.}~\bibnamefont{Pailhés}},
  \bibinfo{author}{\bibfnamefont{A.}~\bibnamefont{Bendounan}},
  \bibinfo{author}{\bibfnamefont{R.}~\bibnamefont{Mottl}},
  \bibinfo{author}{\bibfnamefont{T.}~\bibnamefont{Claesson}},
  \bibinfo{author}{\bibfnamefont{O.}~\bibnamefont{Tjernberg}},
  \bibnamefont{et~al.}, \bibinfo{journal}{New J. Phys.}
  \textbf{\bibinfo{volume}{10}}, \bibinfo{pages}{103016}
  (\bibinfo{year}{2008}).

\bibitem[{\citenamefont{He et~al.}(2011{\natexlab{b}})\citenamefont{He, Zhou,
  Hashimoto, Yoshida, Tanaka, Mo, Sasagawa, Mannella, Meevasana, Yao
  et~al.}}]{He2011b}
\bibinfo{author}{\bibfnamefont{R.-H.} \bibnamefont{He}},
  \bibinfo{author}{\bibfnamefont{X.~J.} \bibnamefont{Zhou}},
  \bibinfo{author}{\bibfnamefont{M.}~\bibnamefont{Hashimoto}},
  \bibinfo{author}{\bibfnamefont{T.}~\bibnamefont{Yoshida}},
  \bibinfo{author}{\bibfnamefont{K.}~\bibnamefont{Tanaka}},
  \bibinfo{author}{\bibfnamefont{S.-K.} \bibnamefont{Mo}},
  \bibinfo{author}{\bibfnamefont{T.}~\bibnamefont{Sasagawa}},
  \bibinfo{author}{\bibfnamefont{N.}~\bibnamefont{Mannella}},
  \bibinfo{author}{\bibfnamefont{W.}~\bibnamefont{Meevasana}},
  \bibinfo{author}{\bibfnamefont{H.}~\bibnamefont{Yao}}, \bibnamefont{et~al.},
  \bibinfo{journal}{New J. of Phys.} \textbf{\bibinfo{volume}{13}},
  \bibinfo{pages}{013031} (\bibinfo{year}{2011}{\natexlab{b}}).

\bibitem[{\citenamefont{Hashimoto et~al.}(2010)\citenamefont{Hashimoto, He,
  Tanaka, Testaud, Meevasana, Moore, Lu, Yao, Yoshida, Eisaki
  et~al.}}]{Hashimoto2010}
\bibinfo{author}{\bibfnamefont{M.}~\bibnamefont{Hashimoto}},
  \bibinfo{author}{\bibfnamefont{R.-H.} \bibnamefont{He}},
  \bibinfo{author}{\bibfnamefont{K.}~\bibnamefont{Tanaka}},
  \bibinfo{author}{\bibfnamefont{J.-P.} \bibnamefont{Testaud}},
  \bibinfo{author}{\bibfnamefont{W.}~\bibnamefont{Meevasana}},
  \bibinfo{author}{\bibfnamefont{R.~G.} \bibnamefont{Moore}},
  \bibinfo{author}{\bibfnamefont{D.}~\bibnamefont{Lu}},
  \bibinfo{author}{\bibfnamefont{H.}~\bibnamefont{Yao}},
  \bibinfo{author}{\bibfnamefont{Y.}~\bibnamefont{Yoshida}},
  \bibinfo{author}{\bibfnamefont{H.}~\bibnamefont{Eisaki}},
  \bibnamefont{et~al.}, \bibinfo{journal}{Nature Physics}
  \textbf{\bibinfo{volume}{6}}, \bibinfo{pages}{414} (\bibinfo{year}{2010}).

\bibitem[{\citenamefont{Razzoli et~al.}(2012)\citenamefont{Razzoli, Drachucki,
  Keren, Radovic, Plumb, Chang, Mesot, and Shi}}]{Razzoli2012}
\bibinfo{author}{\bibfnamefont{E.}~\bibnamefont{Razzoli}},
  \bibinfo{author}{\bibfnamefont{G.}~\bibnamefont{Drachucki}},
  \bibinfo{author}{\bibfnamefont{A.}~\bibnamefont{Keren}},
  \bibinfo{author}{\bibfnamefont{M.}~\bibnamefont{Radovic}},
  \bibinfo{author}{\bibfnamefont{N.~C.} \bibnamefont{Plumb}},
  \bibinfo{author}{\bibfnamefont{J.}~\bibnamefont{Chang}},
  \bibinfo{author}{\bibfnamefont{J.}~\bibnamefont{Mesot}}, \bibnamefont{and}
  \bibinfo{author}{\bibfnamefont{M.}~\bibnamefont{Shi}},
  \bibinfo{journal}{preprint}  (\bibinfo{year}{2012}).

\bibitem[{\citenamefont{Norman et~al.}(2007)\citenamefont{Norman, Kanigel,
  Randeria, Chatterjee, and Campuzano}}]{Norman2007}
\bibinfo{author}{\bibfnamefont{M.~R.} \bibnamefont{Norman}},
  \bibinfo{author}{\bibfnamefont{A.}~\bibnamefont{Kanigel}},
  \bibinfo{author}{\bibfnamefont{M.}~\bibnamefont{Randeria}},
  \bibinfo{author}{\bibfnamefont{U.}~\bibnamefont{Chatterjee}},
  \bibnamefont{and} \bibinfo{author}{\bibfnamefont{J.~C.}
  \bibnamefont{Campuzano}}, \bibinfo{journal}{Phys. Rev. B}
  \textbf{\bibinfo{volume}{76}}, \bibinfo{pages}{174501}
  (\bibinfo{year}{2007}).

\bibitem[{\citenamefont{Eckl et~al.}(2002)\citenamefont{Eckl, Scalapino,
  Arrigoni, and Hanke}}]{Eckl2002}
\bibinfo{author}{\bibfnamefont{T.}~\bibnamefont{Eckl}},
  \bibinfo{author}{\bibfnamefont{D.~J.} \bibnamefont{Scalapino}},
  \bibinfo{author}{\bibfnamefont{E.}~\bibnamefont{Arrigoni}}, \bibnamefont{and}
  \bibinfo{author}{\bibfnamefont{W.}~\bibnamefont{Hanke}},
  \bibinfo{journal}{Phys. Rev. B} \textbf{\bibinfo{volume}{66}},
  \bibinfo{pages}{140510} (\bibinfo{year}{2002}).

\bibitem[{\citenamefont{Mayr et~al.}(2005)\citenamefont{Mayr, Alvarez, \c{S}en,
  and Dagotto}}]{Mayr2005}
\bibinfo{author}{\bibfnamefont{M.}~\bibnamefont{Mayr}},
  \bibinfo{author}{\bibfnamefont{G.}~\bibnamefont{Alvarez}},
  \bibinfo{author}{\bibfnamefont{C.}~\bibnamefont{\c{S}en}}, \bibnamefont{and}
  \bibinfo{author}{\bibfnamefont{E.}~\bibnamefont{Dagotto}},
  \bibinfo{journal}{Phys. Rev. Lett.} \textbf{\bibinfo{volume}{94}},
  \bibinfo{pages}{217001} (\bibinfo{year}{2005}).

\bibitem[{\citenamefont{Valdez-Balderas and Stroud}(2006)}]{ValdezBalderas2006}
\bibinfo{author}{\bibfnamefont{D.}~\bibnamefont{Valdez-Balderas}}
  \bibnamefont{and} \bibinfo{author}{\bibfnamefont{D.}~\bibnamefont{Stroud}},
  \bibinfo{journal}{Phys. Rev. B} \textbf{\bibinfo{volume}{74}},
  \bibinfo{pages}{174506} (\bibinfo{year}{2006}).

\bibitem[{\citenamefont{Zhong et~al.}(2011)\citenamefont{Zhong, Li, and
  Han}}]{Zhong2011}
\bibinfo{author}{\bibfnamefont{Y.-W.} \bibnamefont{Zhong}},
  \bibinfo{author}{\bibfnamefont{T.}~\bibnamefont{Li}}, \bibnamefont{and}
  \bibinfo{author}{\bibfnamefont{Q.}~\bibnamefont{Han}},
  \bibinfo{journal}{Phys. Rev. B} \textbf{\bibinfo{volume}{84}},
  \bibinfo{pages}{024522} (\bibinfo{year}{2011}).

\bibitem[{\citenamefont{Banerjee et~al.}(2011)\citenamefont{Banerjee,
  Ramakrishnan, and Dasgupta}}]{Banerjee2011}
\bibinfo{author}{\bibfnamefont{S.}~\bibnamefont{Banerjee}},
  \bibinfo{author}{\bibfnamefont{T.~V.} \bibnamefont{Ramakrishnan}},
  \bibnamefont{and} \bibinfo{author}{\bibfnamefont{C.}~\bibnamefont{Dasgupta}},
  \bibinfo{journal}{Phys. Rev. B} \textbf{\bibinfo{volume}{84}},
  \bibinfo{pages}{144525} (\bibinfo{year}{2011}).

\bibitem[{\citenamefont{Alvarez et~al.}(2005)\citenamefont{Alvarez, Mayr,
  Moreo, and Dagotto}}]{Alvarez2005}
\bibinfo{author}{\bibfnamefont{G.}~\bibnamefont{Alvarez}},
  \bibinfo{author}{\bibfnamefont{M.}~\bibnamefont{Mayr}},
  \bibinfo{author}{\bibfnamefont{A.}~\bibnamefont{Moreo}}, \bibnamefont{and}
  \bibinfo{author}{\bibfnamefont{E.}~\bibnamefont{Dagotto}},
  \bibinfo{journal}{Phys. Rev. B} \textbf{\bibinfo{volume}{71}},
  \bibinfo{pages}{014514} (\bibinfo{year}{2005}).

\bibitem[{\citenamefont{Alvarez and Dagotto}(2008)}]{Alvarez2008}
\bibinfo{author}{\bibfnamefont{G.}~\bibnamefont{Alvarez}} \bibnamefont{and}
  \bibinfo{author}{\bibfnamefont{E.}~\bibnamefont{Dagotto}},
  \bibinfo{journal}{Phys. Rev. Lett.} \textbf{\bibinfo{volume}{101}},
  \bibinfo{pages}{177001} (\bibinfo{year}{2008}).

\bibitem[{\citenamefont{Vojta et~al.}(2006)\citenamefont{Vojta, Vojta, and
  Kaul}}]{Vojta2006}
\bibinfo{author}{\bibfnamefont{M.}~\bibnamefont{Vojta}},
  \bibinfo{author}{\bibfnamefont{T.}~\bibnamefont{Vojta}}, \bibnamefont{and}
  \bibinfo{author}{\bibfnamefont{R.~K.} \bibnamefont{Kaul}},
  \bibinfo{journal}{Phys. Rev. Lett.} \textbf{\bibinfo{volume}{97}},
  \bibinfo{pages}{097001} (\bibinfo{year}{2006}).

\bibitem[{\citenamefont{Salkola et~al.}(1996)\citenamefont{Salkola, Emery, and
  Kivelson}}]{Salkola1996}
\bibinfo{author}{\bibfnamefont{M.~I.} \bibnamefont{Salkola}},
  \bibinfo{author}{\bibfnamefont{V.~J.} \bibnamefont{Emery}}, \bibnamefont{and}
  \bibinfo{author}{\bibfnamefont{S.~A.} \bibnamefont{Kivelson}},
  \bibinfo{journal}{Phys. Rev. Lett.} \textbf{\bibinfo{volume}{77}},
  \bibinfo{pages}{155} (\bibinfo{year}{1996}).

\bibitem[{\citenamefont{Granath et~al.}(2002)\citenamefont{Granath, Oganesyan,
  Orgad, and Kivelson}}]{Granath2002}
\bibinfo{author}{\bibfnamefont{M.}~\bibnamefont{Granath}},
  \bibinfo{author}{\bibfnamefont{V.}~\bibnamefont{Oganesyan}},
  \bibinfo{author}{\bibfnamefont{D.}~\bibnamefont{Orgad}}, \bibnamefont{and}
  \bibinfo{author}{\bibfnamefont{S.~A.} \bibnamefont{Kivelson}},
  \bibinfo{journal}{Phys. Rev. B} \textbf{\bibinfo{volume}{65}},
  \bibinfo{pages}{184501} (\bibinfo{year}{2002}).

\bibitem[{\citenamefont{Granath and Andersen}(2010)}]{Granath2010}
\bibinfo{author}{\bibfnamefont{M.}~\bibnamefont{Granath}} \bibnamefont{and}
  \bibinfo{author}{\bibfnamefont{B.~M.} \bibnamefont{Andersen}},
  \bibinfo{journal}{Phys. Rev. B} \textbf{\bibinfo{volume}{81}},
  \bibinfo{pages}{024501} (\bibinfo{year}{2010}).

\end{thebibliography}

\end{document}